\documentstyle[11pt,paspconf,epsf]{article}
\begin{document}

\parindent 0.0cm
To appear in Cooling Flows in Galaxies \& Clusters, ASP Conference
Series, N. Soker (ed.)
\parindent=2em
\vskip 1.0cm

\title{``LISTENING" TO CLUSTER COOLING FLOWS: RADIO
EMISSION \& THE CLUSTER ENVIRONMENT}
\author{Jack O. Burns, Chris Loken, Percy G\'omez, Elizabeth Rizza,
Mark Bliton \&  Michael Ledlow}
\affil{Department of Astronomy, New Mexico State University,
Las Cruces, NM 88003 USA}
\author{Frazer N. Owen}
\affil{National Radio Astronomy Observatory, P.O. Box O, Socorro, NM 87801 USA}

\begin{abstract}
The radio continuum properties of galaxy clusters with cooling flows are reviewed along with the
relationship to the X-ray environment.  We find that 60-70\% of cD galaxies in cooling flows are
radio-loud, a much higher fraction than the 14\% found for typical cluster ellipticals. New {\it ROSAT} HRI
observations reveal a variety of interesting correlations and anticorrelations between the X-ray
structure within the inner cooling flows and the radio morphologies.  It appears that the radio plasma
can have a strong effect on the inner structure of cluster cooling cores as  virtually all cooling flow
clusters with central radio sources have non-symmetric X-ray structure.  Numerical simulations of radio
jets in cooling flow atmospheres are presented.  We also discuss the prospects for destroying cooling
flows via cluster-cluster mergers, using new hydro/N-body simulations  which 
incorporate radiative
cooling.
\end{abstract}


\section{Introduction}

The environment at the centers of galaxy cluster cooling flows is exotic and extreme compared to that found near the average galaxy in the Universe.  The intracluster medium (ICM) thermal pressure is a factor of 10-100 times higher than that near other cluster galaxies.  The cooling flow may deposit $\ge$100 M$_{\odot}$/yr onto the central galaxy in such clusters.  Thus, this gaseous environment may be a particularly good one to trigger radio source production, confine the extended radio plasma, and shape the extended radio emission if asymmetries (i.e., pressure gradients) exist in the flow. With such a potentially strong relationship between the environment and the radio
source, radio emission associated with central dominant galaxies may be a useful, generally unappreciated probe of cooling flows.

In $\S$2 of this paper, we begin by presenting a brief overview of cluster
radio sources so that radio galaxies in cooling flows can be placed into a broader relational context.  Next, in $\S$3, the radio continuum properties of supergiant cDs which lie at the centers of clusters are described.  In $\S$4, we discuss the apparent strong relationship between radio structure and the cooling flow environment by comparing VLA and new {\it ROSAT} HRI images.  Then, in $\S$5, numerical simulations of supersonic radio jets in a cooling flow atmosphere are presented in an effort to gain
insight into the radio/X-ray relationships.  In $\S$6, we briefly discuss clusters without cooling flows and present new hydro/N-body simulations which show how cluster-cluster mergers can destroy cooling flows.  Finally, in $\S$7, we summarize our conclusions.

\section{Overview of Cluster Radio Sources}

Much of what follows in this section has been taken from a series of 7 papers which have now been published on the results of the VLA 20-cm survey of $z\le0.09$ Abell clusters
(see e.g., Ledlow \& Owen 1995a,b,1996; Owen \& Ledlow 1997, and references therein).

Cluster radio source morphologies are predominately Fanaroff \& Riley (1974) class I (i.e., low power, edge darkened).  Examples include narrow-angle tailed (NAT) U-shaped sources such as NGC 1265 in Perseus, wide-angle tailed (WAT) sources such as 3C 465 in A2634, and twin-jet sources such as 3C31 (see Owen \& Ledlow 1997 for many examples).  In general, cluster radio source structures are very complex and probably reflect the local ``weather conditions" within the ICM (Burns 1996).  It should be noted that Fanaroff-Riley II sources (i.e., high power, edge-brightened, classical doubles) are rare in nearby rich clusters, but do occur $\approx$7\% of the time in a statistical sample of radio galaxies in Abell clusters (Ledlow \& Owen 1996).

Another intriguing, and poorly understood, property of cluster radio sources is the fact that the radio luminosity function (RLF) is virtually identical for sources inside and outside of rich clusters (Fanti 1984; Ledlow \&  Owen 1996).  This is very surprising since the higher thermal gas pressure inside rich clusters is expected to better confine extended radio sources (and reduce adiabatic expansion losses) and enhance source B-fields, thus creating more luminous radio sources than in radio galaxies outside rich clusters.  However, this picture is clearly too simple.

The probability of a cluster galaxy being a radio source was found to be independent of the global cluster galaxy density (Ledlow \& Owen 1995a).  This, too, seems contrary to the conventional wisdom where (for, at least, the more powerful radio galaxies) the richness of the galaxy neighborhood was thought to be important in triggering radio emission in galaxies (e.g., Heckman et al.~1986).

In an effort to gain some understanding of the complex relationship between extended radio sources and their gaseous environs, we have compared the radio structures with X-ray images of the ICM from {\it{Einstein}} and {\it{ROSAT}} for samples of Abell clusters.  The initial results are intriguing.  We found that larger radio sources in Abell clusters often coincide with clumps of X-ray emission (Burns et al.~1994; Burns 1996) or appear in clusters with significant substructure.  For example, for a sample of 17 clusters with 26 NATs, 88\% have significant X-ray substructure whereas only 23\% of a
comparable sample of radio-quiet Abell clusters have substructure (Bliton et al.
1997).  We also investigated the X-ray emission around a complete sample of 25 $z<0.05$ radio galaxies not cataloged to be in rich clusters; we found that 90\% have X-ray emission similar to the X-ray clumps around radio galaxies in Abell clusters and their optical environments are similar to those in poor clusters (Owen et al.~1996).

From these new data, it appears that the local, not the global, cluster environment is most important in influencing the radio structure.  Similarities in morphologies and RLFs between radio galaxies inside and outside of rich clusters may be caused, at least in part, by similarities in local environments.  It appears that radio galaxies prefer to exist within clusters with complex structures/potentials, which may be evolving as a result of recent cluster/subcluster merging.

\section{Radio Continuum Properties of cD Galaxies}

Supergiant (D \& cD) galaxies are commonly found at the centers of clusters, especially those clusters with cooling flows. These galaxies are exceptional at virtually every wavelength band, including the radio (Ball, Burns, \& Loken 1993).  Whereas a typical elliptical in a cluster has only a 14\% probability of being radio-loud (Ledlow \& Owen 1996), a cD galaxy in a
non-cooling flow cluster has a slightly higher 20\% likelihood of being a radio source.  This increases to 60-70\% for a cD in a cooling flow cluster (Burns 1990; Ledlow et al.~1997). Thus, the large mass and optical brightness of the cD enhances its probability of being a radio source over that of an average elliptical, but its presence in a cooling flow greatly increases
the chances of being radio-loud. Furthermore, the average radio luminosity
($\sim$10$^{42}$ ergs/sec) suggests that the radio plasma energy can be a significant fraction of the thermal energy in the cooling flow if the radio luminosity is only $\sim$1\% of the plasma kinetic energy; thus, the radio sources may be dynamically important influences on the cooling flow as discussed further in the next section.

For a sample of cDs in Abell clusters with {\it{Einstein}} or {\it{ROSAT}} all-sky survey (RASS) images, we do not find any correlation between $L_X$ and the $P_{6cm}$ (6-cm radio power).  However, there may be a weak relationship between \.M and $P_{6cm}$, such that low \.M clusters tend to have low power radio sources (Burns 1990). There are, though,
prominent exceptions such as clusters with WATs which have powerful, extended radio sources yet are rarely ever found in a cooling flow.

A wide variety of radio morphologies are associated with cD galaxies.  Many cDs have typical jet/lobe radio structures (e.g., sources in A1795, A2029, A2199), others have compact ($<$10 kpc) morphologies (e.g., A496), and those not in cooling flows sometimes have wide-angle tails (e.g., A2634).  Interestingly, there is a class of radio morphology that appears to exist only in clusters with cooling flows, which have been dubbed ``amorphous" (Burns 1990).  These sources are typically 100-400 kpc in diameter, have steep radio spectra, strong cores, and diffuse, quasi-spherical structure with little signs of collimated emission such as jets or lobes.  Examples include 3C 84/NGC 1275 in Perseus (Burns et al.~1992) and 3C317 in A2052 (Zhao et al. 1993).  It appears that the cooling flows in such clusters have either disrupted the radio jets or prevented them from forming (Soker \& Sarazin 1988; Loken et al.~1993).

Most of the radio sources associated with cDs in cooling flow clusters lie within the
optical envelopes of the radio galaxies ($\le$50 kpc).  Virtually all these radio
sources have exceptionally steep radio spectra ($\alpha \approx 1-2$, where $S\propto
{\nu}^{-\alpha}$).  This suggests that the sources are confined by the high thermal gas pressure at the
cores of the cooling flows and allowed to spectrally age.  However, cluster radio sources are often
found to have equipartition pressures that are lower than the ICM thermal pressure (e.g., Taylor et
al.~1990; Feretti et al.~1992; Burns et al.~1995).  This may suggest that entrainment of
ICM thermal gas by the radio sources is an important component of the radio plasma
which is not included in the equipartition calculation.  

\section{Relationship Between Radio Structure \& Cooling Flow Environment}

Since the radio sources identified with central galaxies in cluster cooling flows are generally small, high resolution X-ray observations are needed if one is to investigate the relationship between the radio and the ICM/ISM plasmas.  {\it ROSAT} HRI imaging with $\approx$5$\arcsec$ resolution is now
becoming available for a number of cooling flow clusters which allows for detailed comparison with the radio structures.  Although the numbers of observations are still small, some possible trends are beginning to emerge.  We will present our initial impressions here of these trends by dividing these
clusters into 5 categories.

{\it Radio Sources Blowing Bubbles.}  3C 84/NGC 1275 in Perseus is a beautiful example of radio plasma
in the core of the central galaxy apparently evacuating X-ray cavities.  The HRI images presented by
B\"ohringer et al.~(1993) show an anticorrelation between the inner radio structure and the X-ray
emission within a radius of $\approx$1$\arcmin$ of the AGN core.  The X-ray emission associated with
the inner cooling flow is highly asymmetric and appears to have been strongly influenced by the
presence of the radio source.  The radio source has a relatively high total radio power of
$P_{6cm}=10^{25}$ W/Hz.

{\it X-ray Enhancements on the Edges of Radio Sources.}  We have recently analyzed the HRI images of 2 nearby cooling flow clusters, A133 and A2626, which contain steep-spectrum radio sources.  These are
only moderately powerful radio sources at 6-cm ($P_{6cm}=4\times10^{22}$ W/Hz for A133 and $P_{6cm}=5\times10^{22}$ W/Hz for A2626). After subtracting out circular model-fits to the X-ray isophotes, we find X-ray excesses which nicely correlate with the steep-spectrum extended radio
emission.  The example of A133 is shown in Fig. 1.  We find that inverse Compton scattering of microwave background photons by the relativistic electrons in the radio plasma predicts X-ray emission which is $\sim$100 times less than what is observed.  So, it appears that the X-ray excess is more likely to be
hot gas (possibly ICM/ISM gas compressed by the bow shock of the radio source, see below); we estimate the X-ray ``clouds'' to have masses of $10^5-10^6$ M$_{\odot}$.

\begin{figure}[tbh]
\epsfxsize=4.2in \epsfbox{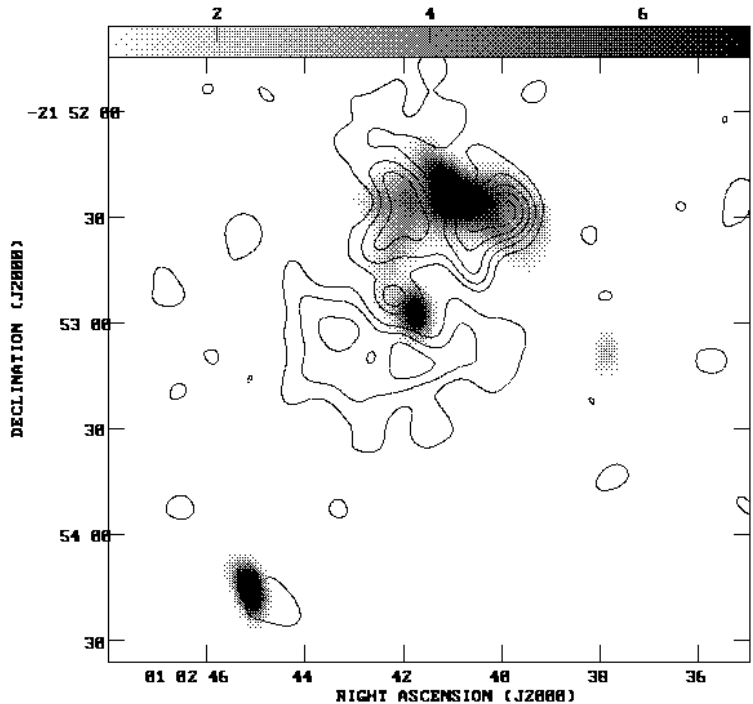}
\caption{Grey-scale image of the radio source in Abell 133 overlaid with
contours representing the residual X-ray emission after subtraction
of a circularly-symmetric model of the cluster emission. Contours to the south
of the radio core show a negative hole.  Note the strong
spatial correlation between the positive excess X-ray emission to the north
of the radio core ($\approx$10$\sigma$) and the radio-emitting plasma.}
\end{figure}

{\it Bubbles and X-ray Enhancements.}  Cygnus A is an example of a cooling flow cluster which shows
both X-ray enhancements near the outer hot spots of the radio lobes as well as X-ray holes within the
lobes closer to the nucleus (Carilli, Perley, \& Harris 1994).  Clarke \&  Harris (1996) used hydro simulations to
explain these observations -- the enhancements are compressed gas near the source bow shock and the X-ray holes are produced by gas which has been evacuated by the radio plasma propagating through the
cooling flow.  Cygnus A is, of course, one of the most powerful classical double sources,
$P_{6cm}=2\times10^{27}$ W/Hz.

{\it X-ray Extensions Between the Radio Tails/Lobes.}  We have also begun to see a few interesting
examples of correlations between asymmetries in the inner cooling flows and bent radio structures. 
A2597 (Sarazin et al.~1995; $P_{6cm}=5\times10^{24}$ W/Hz) and A2199/3C 338 (Owen \& Eilek 1997;
$P_{6cm}=8\times10^{23}$ W/Hz) show excess X-ray emission between the radio tails or lobes of these
small linear size sources.  Both the X-ray asymmetries and the bent radio sources suggest that there
are important pressure gradients within the inner cooling flow regions of these cD galaxies.

{\it Alignment of Radio Source with Dust \& Emission-line Gas.} Finally, a recent, spectacular example
of an unanticipated correlation between inner dust lanes, outer emission-line gas, and extended radio
features has been revealed for A1795 using new HST imaging (Pinkney et al.~1996; McNamara et al.~1996). 
Clearly, these phenomena are all related in a yet-to-be-determined manner, making the inner cooling
flow environment far more complex than previously believed.

In summary, the new HRI and HST images for cooling flows with central radio sources clearly show them
to be highly nonspherical.  Both the radio luminosities and now the X-ray/radio correlations suggest
that the radio plasma has an important impact on the central structure of the cooling flow.  The
highest radio power sources (e.g., Cygnus A \& 3C 84) are capable of blowing bubbles or evacuating
cavities in the inner cooling flow region, whereas lower radio power sources can produce X-ray
excesses which coincide with the outer portions of the radio structures.

\section{Numerical Simulations of Radio Jets in a Cluster Cooling Flow}

In an effort to understand the complex relationship between extended radio sources and the cooling flow environment, Loken et al.~(1993) performed the first numerical simulations of supersonic radio jets propagating through a realistic cooling flow atmosphere. These 2-D hydrodynamics simulations
launched 4 jets of differing Mach numbers (3, 6, 12, and 50 as  measured with respect to the jet internal sound speed) into a steady-state cooling flow.  One example of the end state of a 1.3 kpc long,  M=12 jet is shown in Fig. 2.

\begin{figure}[h]
  \plotone{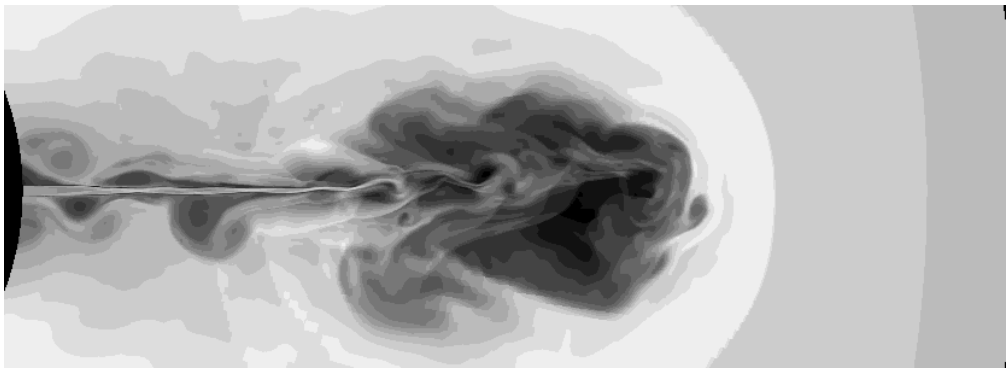}
\caption{Grey-scale image of gas density at a late stage in the evolution of
a perturbed, 2D, Mach 12 jet in a cooling flow atmosphere. The jet
has disrupted and is forming a large, low-density (dark) turbulent lobe. 
The region shown is 200 jet radii long. See Loken et al.~(1993) for more details.
}
\end{figure}

For a laminar jet, one can show using simple ram  pressure balance that the morphology and length of a
radio jet should be a strong function of jet and cooling inflow  Mach numbers (Loken et al.~1993). 
Clearly, lower Mach number jets (lower thrust) should not travel as far into a high inflow velocity
cooling flow as would a high Mach number jet in a low  velocity cooling flow.  However, this naive
calculation does not consider the fact that fluid shear or Kelvin-Helmholtz instabilities play a
dominant role as the jet fluid attempts to propagate through the inflowing cooling cluster gas. The
jets become unstable, ultimately disrupt, and stagnate as shown in Fig. 2.  The disruption length of a
jet is once again a strong function of jet and cooling flow  Mach numbers.  Thus, a M=3 jet stagnates
close to the radio core, whereas a high  Mach number jet can ``blow through" the central cooling flow
with little effect. This may be why low luminosity radio sources in central cooling flow galaxies are
generally small (low thrust jets), whereas a powerful source such as Cygnus A is able to grow to a
diameter of $\approx$200 kpc.

For the M=12 jet simulation shown in Fig. 2, we have computed the projected X-ray brightness
distribution in the {\it ROSAT} band produced by the hot ISM/ICM surrounding the jet.  We find that the
X-ray brightness around the jet is $\approx$10-20\% higher than would be expected from the cooling
flow alone.  It appears that the jet bow shock has compressed and heated the gas in a shell around the
radio source.  Since this jet is within the inner $\approx$2 kpc of the cooling flow center where the
gas temperature is only $\approx$10$^6$ K, the bow shock heating produces a significant enhancement in
X-ray emissivity in the {\it ROSAT} band.  Thus, X-ray excesses on the edges of radio components such
as that seen in Fig. 1 may be understood as shocked, compressed gas produced as the radio plasma
propagates through the cores of cooling flows.

\section{Cluster-Cluster Mergers \& the Death of Cooling Flows}

Although many (possibly most) galaxy clusters contain cooling flows, there are also a number of
clusters whose temperature and X-ray surface brightness profiles suggest a more isothermal gas at the
cluster core.  Two good examples of such classes of non-cooling flow clusters are those which contain
wide-angle tailed radio galaxies (e.g., G\'omez et al.~1997) and radio halos (Burns et al.~1995).
Both of these classes of clusters show strong evidence for substructure in  X-ray images (e.g., Briel
et al.~1991 for the radio halo cluster A2256) and in cluster velocity distributions (e.g., Pinkney et
al.~1993 for the WAT cluster A2634).  These data have been used to model these clusters as candidates
for cluster-cluster mergers (e.g., Roettiger et al.~1995; 1997).

Do such mergers also destroy previously existing cooling flows? To address this question, we performed
2-D simulations of a  merger between two idealized spherical clusters with a mass ratio of 4:1 using
our hydro/N-body code (see Roettiger et al.~1997 for details).  In the larger cluster with $10^{15}$
M$_{\odot}$, $T$=11.7 keV, and $r_{core}$=250 kpc, a steady-state cooling flow with mass dropout was
initially evolved in 1-D and then placed within the 2-D grid.  The second cluster was initially
isothermal with $T$=6.6 keV and $r_{core}$=157 kpc.  The two clusters were allowed to fall together
under the influence of gravity.  In this simulation, cooling via bremsstrahlung and line radiation was
activated using the cooling function described in 
Westbury \& Henriksen (1992).  Three epochs during the cluster-cluster
collision are displayed in Fig. 3.  The net effect of the collision is strong shock heating of the
cooling flow core which causes it to expand, the temperature increases in the core from  1.5 keV to 15
keV, and the cooling time increases to $>20$ Gyrs in the final epoch. Thus, for the particular
parameters in this simulation, the cooling flow is effectively destroyed and will not become re-established within a Hubble time.

\begin{figure}[bt]
  \plotone{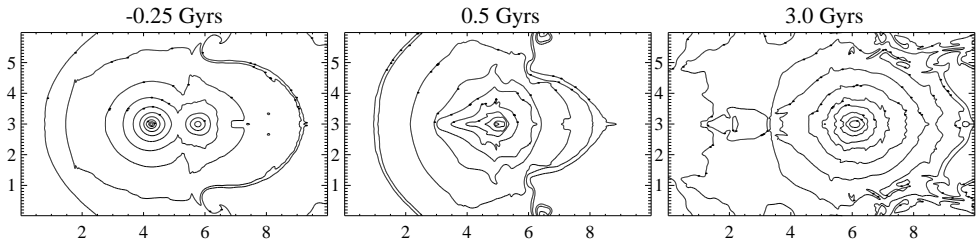}
\caption{Contours of gas density at 3 stages during the evolution of
a cluster merger. The subcluster falls from the right towards
the 4-times more massive, cooling flow cluster (first panel). Shortly after
the cores overlap, there are elongations in the gas distribution both
perpendicular and parallel to the merger axis (middle panel). 3 Gyrs after
the merger (final panel), the main cluster is relatively spherical but
there is no evidence for a cooling flow (note less compact core than in
first panel). Axes are labelled in Mpc; times
are relative to core-crossing. 
}
\end{figure}
It is probably premature to generalize this result to the life history of most clusters.  This is a ``hard", head-on collision (high momenta) between two massive clusters which may be relatively rare in the Universe.  Although cosmological large-scale structure simulations do find that clusters grow
hierarchically through mergers and accretion of gas and dark matter along filaments (e.g., Bryan et al.~1994; Loken et al.~1997), this process may generally be softer and not destroy a cooling flow in all cases.  Such a
picture must be true to explain both the preponderance of substructure and cooling flows in nearby clusters.  We are in the process of performing additional simulations with higher mass ratios to explore this issue further.

\section{Summary}

There appears to be a very strong correlation between radio emission and the presence of cooling flows
around central galaxies in clusters.  Nearly three-quarters of cD galaxies at the centers of cooling
flows are radio-loud whereas only 14\% of average cluster ellipticals have radio emission in surveys of
Abell clusters.  Radio sources associated with dominant galaxies in cooling flows tend to be small
($<50$ kpc in diameter) and have steep radio spectra ($\alpha$=1-2).  A class of radio source, termed
amorphous, which have little collimated emission (i.e., jets or lobes) and instead have a core-halo
(steep spectrum halo) morphology are found to exist only in cooling flows.

New {\it ROSAT} HRI images reveal strong interactions between radio plasma and the centers of cooling flows.  ``Bubbles" or evacuated cavities are seen in the X-ray emission in cooling flows around the most powerful radio sources such as 3C 84 in the Perseus cluster and near the classical double Cygnus
A.  In other cases such as A133 and A2626, enhancements of X-ray emission are seen on the edges of the extended radio emission, possibly produced by compression and heating of the ISM/ICM by the radio
source bow shock as suggested by numerical simulations.  Finally, in yet other cases such as A2199 and A2597, we observe alignments between bent radio jets/tails and extended central X-ray emission.  In virtually every cooling flow cluster with a central radio source, the X-ray emission is observed to be
highly asymmetric.  Such properties need to be considered in modeling cooling flows (see e.g., Garasi et al.~1997).

New hydro/N-body simulations suggest that cooling flows can be destroyed by low mass ratio cluster-cluster collisions.  Such collisions are consistent with X-ray and optical observations of clusters
which contain WAT radio galaxies or cluster halos -- neither of which possess cooling flows.  Further
simulations will be needed to reconcile the observed abundance of both cooling flow clusters and
substructure (i.e., merging) within clusters.

\acknowledgments
We thank our colleagues who have contributed to this work including  Wolfgang Voges, Anatoly Klypin,
Kurt Roettiger, Neal Miller, Rusty Ball, and Jason Pinkney.  This research was supported by grants from
the U.S. National Science Foundation (AST-9317596) and NASA (NAGW-3152). NRAO is operated by
Associated Universities, Inc., under cooperative agreement with the National Science Foundation.

\end{document}